\documentclass[]{aastex631}

	\usepackage{hyperref}

\begin{document}

		\title{Revealing the Fate of Exoplanet Systems: \\
		Asteroseismic Identification of Host Star in the Red Clump or Red Giant Branch}

		\correspondingauthor{Sheng-Bang Qian}
		\email{qiansb@ynu.edu.cn}

		\author[0000-0002-8276-5634]{Wen-Xu Lin}
		\affiliation{Yunnan Observatories, Chinese Academy of Sciences, Kunming 650216, People's Republic of China}
		\affiliation{Department of Astronomy, School of Physics and Astronomy, Yunnan University, Kunming 650091, P. R. China}
		\affiliation{Key Laboratory of Astroparticle Physics of Yunnan Province, Yunnan University, Kunming 650091, P. R. China}
		\affiliation{University of Chinese Academy of Sciences, No.1 Yanqihu East Rd, Huairou District, Beijing, 101408, People's Republic of China}

		\author[0000-0002-5995-0794]{Sheng-Bang Qian}
		\affiliation{Department of Astronomy, School of Physics and Astronomy, Yunnan University, Kunming 650091, P. R. China}
		\affiliation{Key Laboratory of Astroparticle Physics of Yunnan Province, Yunnan University, Kunming 650091, P. R. China}

		\author[0000-0002-0796-7009]{Li-Ying Zhu}
		\affiliation{Yunnan Observatories, Chinese Academy of Sciences, Kunming 650216, People's Republic of China}
		\affiliation{University of Chinese Academy of Sciences, No.1 Yanqihu East Rd, Huairou District, Beijing, 101408, People's Republic of China}

	\begin{abstract}

		Determining the evolutionary stage of stars is crucial for understanding the evolution of exoplanetary systems. In this context, Red Giant Branch (RGB) and Red Clump (RC) stars, stages in the later evolution of stars situated before and after the helium flash, harbor critical clues to unveiling the evolution of planets. The first step in revealing these clues is to confirm the evolutionary stage of the host stars through asteroseismology. However, up to now, host stars confirmed to be RGB or RC stars are extremely rare. In this investigation, we present a comprehensive asteroseismic analysis of two evolved stars, HD 120084 and HD 29399, known to harbor exoplanets, using data from the Transiting Exoplanet Survey Satellite (TESS). We have discovered for the first time that HD 120084 is a Red Clump star in the helium-core burning phase, and confirmed that HD 29399 is a Red Giant Branch star in the hydrogen-shell burning phase. Through the precise measurement of asteroseismic parameters such as $\nu_{max}$, $\Delta\nu$ and $\Delta\Pi_{1}$ we have determined the evolutionary states of these stars and derived their fundamental stellar parameters. 
		The significance of this study lies in the application of automated techniques to measure asymptotic period spacings in red giants, which provides critical insights into the evolutionary outcomes of exoplanet systems. We demonstrate that asteroseismology is a potent tool for probing the internal structures of stars, thereby offering a window into the past and future dynamics of planetary orbits. The presence of a long-period giant planet orbiting HD 120084, in particular, raises intriguing questions about the potential engulfment of inner planets during the host star's expansion, a hypothesis that warrants further investigation.

	\end{abstract}


	\section{Introduction} \label{sec:intro}

	In recent years, the development of space telescopes has ushered in an era of big data in several directions within the field of astronomy, such as exoplanet science, asteroseismology, and binary star studies \citep{chaplin2013asteroseismology, 2019LRSP...16....4G, bowman2020asteroseismology, southworth2021space, 2021FrASS...7..102J}. After many years of exploration, the study of exoplanets has moved beyond merely discovering new exotic worlds. It now involves more statistical work \citep{winn2015occurrence, mulders2018planet}, evolution and development studies \citep{jontof2019compositional, hon2023close}, as well as detailed research on planetary atmospheres or magnetic fields \citep{guo2024characterization, giacobbe2021five}.

	Studying the planets that orbit around stars in their late evolutionary stages is always fascinating because the evolution of a star after it leaves the main sequence directly impacts the fate of its planetary companions. These planets can often provide us with clues about the future fate of Earth. Because of the limitations of observational methods, the current ways astronomers discover planets are mostly confined to the study of host stars. For example, they explore the gravitational perturbation of planets on stars by studying the changes in radial velocity of the host stars and detect signals of planetary transits by observing the minor changes in the host star's light \citep{mayor1995jupiter, konacki2003extrasolar}. Asteroseismology, as a discipline that studies the internal structure of stars by analyzing their oscillations, also serves as an important window for researching exoplanets and evolved host stars. \citep{2018arXiv180402214L}. 

	In the asteroseismic study of stars in the late stages of evolution, two global asteroseismic parameters derived from the radial modes are very important: the frequency at maximum power, $\nu_{\max}$, and the large frequency separation $\Delta\nu$, which are the frequency corresponding to the maximum power of the star’s solar-like oscillation modes and the frequency difference between adjacent oscillation modes, respectively. If the effective temperature of a star can be obtained through spectroscopic observations, it allows for the direct calculation of the stellar radii, masses, surface gravity and luminosity \citep{chaplin2013asteroseismology}. The oscillation spectra of Evolved stars also exhibit mixed mode, which were identified in red giants by \cite{beck2011kepler}. The most classic example is the dipole mixed modes, whose amplitudes are larger than those of other mixed modes. The $l = 1$ mixed modes results from the coupling of the dipole pressure modes and the dipole gravity modes \citep{bedding2010solar}. Since they possess characteristics of gravity modes (g-modes), these mixed modes provide a window for studying the structure of the stellar cores \citep{stello2013asteroseismic}. Dipole mixed modes were used to distinguish core-helium burning giants (red clump stars) from hydrogen-shell burning giants (red giant branch star) \citep{bedding2011gravity, stello2013asteroseismic}. The gravity modes are distributed at equal intervals in the period domain, and the interval value changes with the size of the burning core. The observable mixed modes are distributed at different intervals, denoted as $\Delta P$, in the period domain for red giant branch (RGB) stars with hydrogen shell burning and red clump (RC) giants with helium core burning, showing different patterns. In the research conducted by \cite{stello2013asteroseismic}, they analyzed and categorized more than 13,000 red giants observed by Kepler. They found that for RGB stars, the median($\Delta P$) falls between 30 to 70 seconds, whereas for red clump stars, the median($\Delta P$) is between 200 to 270 seconds. Later on, \cite{vrard2016period} developed a method to calculate the period spacing of dipole gravity modes without identifying the dipole mixed modes. This method can be directly applied to the power spectrum and performs better at low signal-to-noise ratios compared to identifying individual mixed modes. This method can more conveniently and effectively determine the evolutionary stage of red giants.

	In this work, we performed a detailed asteroseismic analysis on two host stars, HD 120084 and HD 29399, which are in the later stages of evolution. By obtaining asteroseismic parameters, we not only recalculated the physical parameters of the stars but also adjusted the parameters of the accompanying exoplanets using the method described by \citet{2024AJ....168...27L}. This highlights the potential of the intersection between asteroseismology and exoplanetary science.. Furthermore, by calculating the asymptotic period spacing of dipole gravity modes, we determined that HD 120084 is a red clump giant star burning helium in its core, while HD 29399 is confirmed to be a red giant branch star burning hydrogen in its shell. Given that the current samples of host stars with determined evolutionary states are extremely rare \citep{hon2023close, campante2019tess}, work of this nature is incredibly important. Future further observations and in-depth studies of these systems will have a profound impact on research into the evolution of planetary systems.

	\section{Asteroseismic analysis} \label{sec:Asteroseismic analysis}

	\subsection{Light Curve Preprocessing and Estimation of Astroseismic Parameters} \label{subsec:preprocess}

	At the outset, we utilized the data on subgiants from \citet{straivzys1981fundamental} to plot a calibration line on the Hertzsprung-Russell diagram correlating effective temperature and surface gravity. We selected those host stars \citep{STELLARHOSTS} that lie above the calibration line (noting that the surface gravity axis is inverted), with their latest effective temperature and surface gravity data sourced from the \citet[see \ref{tab:paras_from_paper}]{ps}. In addition to the two host stars in this work, HD 120084 and HD 29399, we have also found dozens of host stars with typical solar-like oscillation light curves, i.e., their power spectra have significant Gaussian-distributed power excess. Among these host stars, some have had their asteroseismic analyses published, such as 8 UMi \citep{hon2023close}, HD 212771, and HD 203949 \citep{2019ApJ...885...31C}. There are also contents from our previous work, such as HD 5608, HD 33844, 7 CMa, and HIP 67851 \citep{2024AJ....168...27L}. However, for the vast majority of host stars, when we tried to analyze their mixed modes using the methods introduced in this work, we found that due to the lack of signal-to-noise ratio, we could not determine their evolutionary states.
	Among the identified stars, two in particular, HD 120084 (TIC 284181945) and HD 29399 (TIC 38828538), have asteroseismic information that allows for a direct assessment of their evolutionary status.
	
	Python package $lightkurve$ \citep{2018ascl.soft12013L} was used to download  all available 2-minute cadence light curves \footnote[1]{For the two host stars involved in this work, the number of sectors for the 2-minute cadence light curves is significantly higher than other cadences (e.g., 10-minute, 30-minute), making the 2-minute cadence data the preferred choice.} of HD 120084 and HD 29399 from the Mikulski Archive for Space Telescope (MAST, https://mast.stsci.edu/portal/Mashup/Clients/Mast/Portal.html). These light curves were extracted and de-trended by the TESS Science Processing Operations Center (SPOC) pipeline \citep{jenkins2016tess}.
	Next, we applied the Lomb–Scargle method \citep{lomb1976least, scargle1982studies} to analyze the light curve and obtain the power spectrum, then the \texttt{pySYD} package \citep{2022JOSS....7.3331C} could be used to obtain an estimate of the values of the large frequency separation ($\Delta \nu$) and frequency at maximum power ($\nu _{max}$). To obtain the signal-to-noise ratio (SNR) diagram, we removed the estimated background noise from the power spectrum. The background noise was calculated by smoothing filter with a width of $log_{10}(0.01 \mu Hz)$. The SNR data could be input into the \texttt{PBjam}\footnote[2]{https://github.com/grd349/PBjam} \citep{2021AJ....161...62N} package for analysis to determine the frequencies of the observed radial, $\nu_{n,0}$, and quadrupolar, $\nu_{n,2}$, oscillation modes of host stars.
	Appropriate prior distributions of global asteroseismic parameters and effective temperature are necessary for \texttt{PBjam} package, as they can constrain the parameter space and improve the reliability of the results. Therefore, we can use the global asteroseismic parameters provided by \texttt{pySYD} and the effective temperatures given in the literature, along with reasonable error ranges, as prior distributions.

	\subsection{Deriving Stellar Fundamental Parameters from Asteroseismic Parameters} \label{subsec:stellarparameters}

	After calculating $\nu _{max}$ and $\Delta \nu$, along with the $T_{eff}$ derived from spectral data, the stellar mass, radius, and other physical parameters can be estimated based on asteroseismic scaling relations \citep{1995A&A...293...87K, chaplin2013asteroseismology}. However, for red giants, their evolutionary state and metallicity can affect these scaling relations. \citet{sharma2016stellar} incorporated correction factors into the scaling relations, 
	\begin{equation}\label{equation:1}
		(\frac{R}{R_{\odot } } ) \simeq (\frac{\nu_{max}}{f_{\nu_{max}}\nu_{max,\odot} } )(\frac{ \Delta\nu }{f_{\Delta\nu}\Delta\nu_{\odot} } )^{-2} (\frac{T_{eff}}{T_{eff,\odot} })^{0.5},
	\end{equation}
	\begin{equation}\label{equation:2}
		(\frac{M}{M_{\odot } } ) \simeq (\frac{\nu_{max}}{f_{\nu_{max}}\nu_{max,\odot} } )^{3}(\frac{\Delta\nu}{f_{\Delta\nu}\Delta\nu_{\odot} } )^{-4} (\frac{T_{eff}}{T_{eff,\odot} })^{1.5},
	\end{equation}
	\begin{equation}\label{equation:3}
		\Delta\nu \simeq f_{\Delta\nu}(\frac{\rho}{\rho_{\odot}})^{0.5},
	\end{equation}
	\begin{equation}\label{equation:4}
		\frac{\nu_{max}}{\nu_{max,\odot}} \simeq f_{\nu_{max}}\frac{g}{g_{\odot}}(\frac{T_{eff}}{T_{eff,\odot} })^{-0.5}.
	\end{equation}

	The correction factors, $f_{\nu_{max}}$ and $f_{\Delta\nu}$, are related to stellar mass, metallicity, evolutionary state and effective temperature \citep{sharma2016stellar}. The specific correction methods (\texttt{Asfgrid}) can be found at this website\footnote[3]{http://www.physics.usyd.edu.au/k2gap/Asfgrid/} \citep{sharma2016stellar, stello2023extension}.
	When using the \texttt{Asfgrid} code to calculate stellar mass and radius, the required inputs are global asteroseismic parameters, effective temperature, metallicity (see Appendix Table \ref{tab:paras_from_paper}), and evolutionary state (which can be determined in subsequent analysis steps).
	The solar effective temperature, $T _{\odot } $, is equal to 5772.0 $\pm$ 0.8 K \citep{prvsa2016nominal}.
	The solar $\nu _{max}$ is equal to 3141 $\pm$ 12 $\mu$Hz, and the solar $\Delta \nu$ is equal to 134.98 $\pm$ 0.04 $\mu$Hz \citep{andersen2019oscillations}.

	\subsection{Automated measurement of period spacings in red giants} \label{subsec:periodspacing}

	The oscillation spectrum of red giants exhibits mixed modes, and the characteristic of the dipole mixed modes allow us to distinguish between helium-core burning giants (Red Clump, RC) and hydrogen shell-burning giants (Red Giant Branch, RGB) \citep{bedding2011gravity}. Due to the different physical conditions such as radiative cores and chemical composition gradients between RC and RGB stars, the typical value of the asymptotic period spacing of dipole gravity modes, denoted as $\Delta\Pi_{1}$, is different \citep{mosser2012probing}. The typical value of $\Delta\Pi_{1}$ for RC stars is between 280 and 320 seconds, while the typical value of $\Delta\Pi_{1}$ for RGB stars is between 65 and 90 seconds \citep{vrard2016period}. 

	In actual observations, what we can obtain from light curves is the period spacing of consecutive mixed modes, namely the characteristic feature of the dipole mixed modes, which is denoted as $\Delta P$. However, the observable period spacing of $\Delta P$ is not a constant value. Its relationship with the period spacing of dipole gravity modes, $\Delta\Pi_{1}$, can be expressed by the following equation \citep{2015A&A...580A..96D, 2015A&A...584A..50M, vrard2016period},
	\begin{equation}\label{equation:5}
		\Delta P = \zeta  \Delta\Pi_{1} ,
	\end{equation}

	\begin{equation}\label{equation:6}
		\zeta=\left[1+\frac{1}{q}\frac{\nu^{2} \Delta\Pi_{1}}{\Delta\nu(n_{p})}\frac{\cos^{2}\pi\frac{1}{\Delta\Pi_{1}} \left(\frac{1}{\nu}-\frac{1}{\nu_{g}}\right)}{\cos^{2}\pi\frac{\nu-\nu_{p}}{\Delta\nu(n_{p})}}   \right]^{-1} .
	\end{equation}
	where $\nu_{p}$ and $\nu_{g}$ are the asymptotic frequencies of pure pressure and gravity modes, $\Delta\nu(n_{p})$ is the frequency difference between two consecutive pure pressure radial modes with radial orders $n_{p}$ and $n_{p} +1$, and q is the coupling parameter between the pressure and gravity-wave patterns \citep{1979nos..book.....U, 1989nos..book.....U}. In actual calculations, one method to calculate $\Delta\Pi_{1}$ is to perform a stretching transformation on the power spectrum, we need to use the $\zeta$ function to perform an extension transformation on the SNR spectrum in the period domain, 
	
	\begin{equation}\label{equation:7}
		\mathrm{d}\tau =\frac{1}{\zeta} \frac{\mathrm{d}\nu}{\nu^{2}}  .
	\end{equation}
	
	Using Equation \ref{equation:7}, we can obtaining a new spectrum, $P(\tau)$. The coefficient $\zeta$ in Equation \ref*{equation:7} requires appropriate values of $\Delta\Pi_{1}$ and $q$ for the mixed modes in $P(\tau)$ to be evenly spaced, with the spacing equal to $\Delta\Pi_{1}$. The specific application of this step will be detailed in the subsequent data analysis section. However, the $\zeta$ function is not continuous. Based on the characteristic that the $\zeta$ function is not sensitive to small changes in $\Delta\Pi_{1}$ values, it is generally necessary to use multiple different $\Delta\Pi_{1}$ values to interpolate a specific $\zeta$ function to make it continuous. Computing the zeta function is not easy, but there is a simplification \citep{cunha2019analytical},
	\begin{equation}\label{equation:8}
		\zeta = \left [1+\frac{\nu^{2}}{\nu_{g}^{2}} \frac{q}{\nu_{p}} \left [ \sin^{2}(\frac{\nu - \nu_{a,n}}{\nu_{p}} ) +q^{2} \cos^{2}(\frac{\nu - \nu_{a,n}}{\nu_{p}})\right ]^{-1}   \right ] ^{-1} ,
	\end{equation}
	$\nu_{a,n}$ is the frequency of what would be the pure acoustic mode of (pressure) radial order n, in the absence of mode coupling. 

	After identifying the $l=0$ and $l=2$ pressure modes using the \texttt{PBjam} package, we removed the spectral information between $\nu_{n,2} - 0.1\Delta\nu$ and $\nu_{n,0} + 0.1\Delta\nu$ in the frequency domain. This step aims to filter out the dipole mixed modes we need. Additionally, in high SNR data, $\nu_{n,3}$ modes may appear. Their presence does not affect the measurement of $\Delta\Pi_{1}$ in the actual data analysis but does impact the error calculation of $\Delta\Pi_{1}$. However, if removal is necessary, their positions need to be visually identified in the echelle diagram, and the range of spectral information removal should be finely adjusted.

	By performing a Fourier transform on $P(\tau)$ and converting it to the period domain, a significant peak or group of peaks can be obtained at a position where the periodic coordinate equals the $\Delta\Pi_{1}$. From the research results of Vrard et al., it can be seen that changes in the $\Delta\Pi_{1}$ value over a large range (e.g., 70s or 300s) significantly affect the $\zeta$ function, while changes over a small range (e.g., 70s or 75s) have less impact. The variation in the $q$ value has a similar impact on the $\zeta$ function \citep{vrard2016period}. Therefore, in the actual calculations, we first fix the values of $\Delta\Pi_{1}$ and $q$ at (75s, 0.15) or (300s, 0.3) (typical values for RGB and RC stars, respectively). These values are then used in the data processing procedure to obtain the $\Delta\Pi_{1}$ value, which serves as an initial estimate. Next, we center the obtained $\Delta\Pi_{1}$ value and set 201 $\Delta\Pi_{1}$ values with a resolution of 0.1s. Additionally, we center around the typical $q$ value (0.15 or 0.3) and set 21 $q$ values with a resolution of 0.01 as the second dimension. This forms a 2-dimensional grid of input values, which is then used in the aforementioned method to obtain the Fourier transform spectrum of $P(\tau)$. When the input and output values of $\Delta\Pi_{1}$ are closest, and the amplitude of the Fourier transform spectrum at the $\Delta\Pi_{1}$ value is the highest, the input value of $q$ can be considered accurate, and the output value of $\Delta\Pi_{1}$ is accurate.

	After the stretch, the uncertainty (resolution of the spectrum) introduced into the $\Delta\Pi_{1}$ is \citep{vrard2016period}
	\begin{equation}\label{equation:9}
		\delta(\Delta\Pi_{1})_{res} = \nu_{max} \Delta\Pi_{1}^{2} .
	\end{equation}

	For higher density of mixed modes data, calculating the spectrum of $P(\tau)$ results in oversampling. Thus, the error in calculating $\Delta\Pi_{1}$ can be determined using the following equation \citep{vrard2016period},
	\begin{equation}\label{equation:10}
		\delta(\Delta\Pi_{1})_{over} \simeq \frac{1.6}{A}\delta(\Delta\Pi_{1})_{res} ,
	\end{equation}
	where A is the amplitude corresponding to the $\Delta\Pi_{1}$ value in the spectrum of $P(\tau)$.

	\begin{figure}[ht!]

	\includegraphics[width=0.75\linewidth]{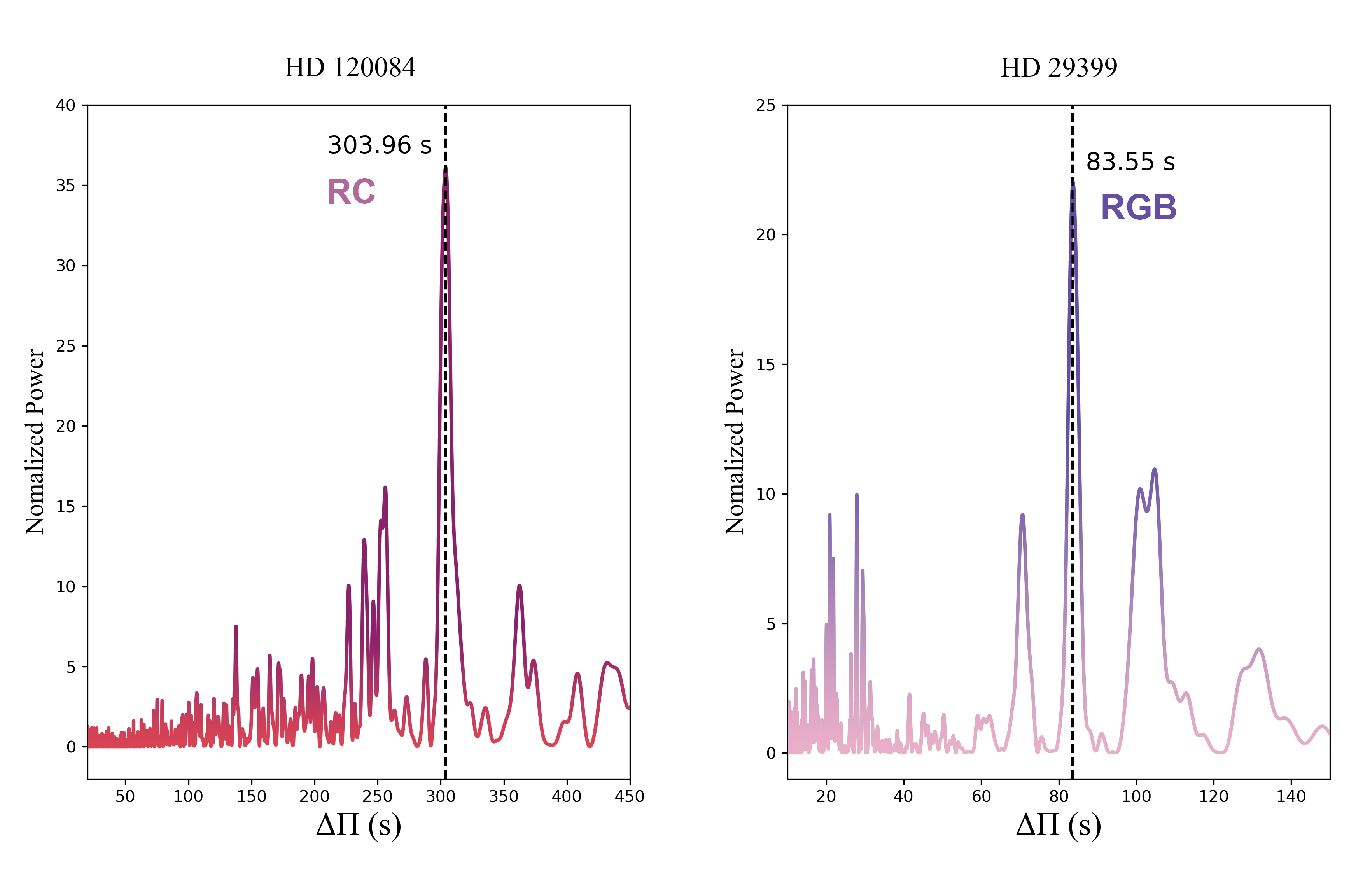}
	\centering
	\caption{\textbf{Left}: the power spectrum of $P(\tau)$  for the host star HD 120084, $\Delta\Pi_{1}$ for this star is 303.96 s. \textbf{Right}: the power spectrum of $P(\tau)$ for the host star HD 29399, $\Delta\Pi_{1}$ for this star is 83.55 s.
	\label{fig:deltapi}}

	\end{figure}

	\begin{splitdeluxetable*}{cccccccBcccccccccc}
	\renewcommand{\arraystretch}{1.5}
	\tabletypesize{\scriptsize}
	\tablewidth{0pt} 
	%\tablenum{1}
	\tablecaption{Parameters of system HD 120084 and HD 29399 \label{tab:para}}
	\tablehead{
	\colhead{}  &
	\multicolumn{6}{c}{Parameters of Host Star} & \multicolumn{5}{c}{Global Seismic Parameters} &\colhead{} &\colhead{} & \multicolumn{3}{c}{Parameters of Planet}\\
	\cline{2-7}
	\cline{8-12}
	\cline{15-17}
	\colhead{Host Name} & \colhead{$T_{eff}$} & \colhead{$L_{\ast }$}  & \colhead{$M_{\ast }$} & \colhead{$R_{\ast }$}  & \colhead{$\log_{}{g} $} & \colhead{Evolutionary Stage} & \colhead{$\nu_{max}$} &  \colhead{$\Delta\nu$} &  \colhead{$\Delta\Pi_{1}$} &  \colhead{$\delta\Delta\Pi_{1}$} &  \colhead{q} & \colhead{} & 
	\colhead{Planet Name}  & \colhead{$M_{p}\sin i$} & \colhead{$M_{p}\sin i$}  &  \colhead{a}\\
	\colhead{} & \colhead{K} & \colhead{$\log_{10}{L_{\odot }} $} & \colhead{$M_{\odot } $} & \colhead{$R_{\odot } $}   & \colhead{$\log_{10}{\mathrm{(cm/s^2)}} $} &  \colhead{} & \colhead{$\mu Hz$} &  \colhead{$\mu Hz$} &  \colhead{s} &  \colhead{s} &  \colhead{}  &  \colhead{} &  \colhead{} & \colhead{$M_{\oplus }$} & \colhead{$M_{Jupiter}$}  & \colhead{AU}
	} 
	\startdata 
	{HD 120084}&4892$\pm  22$ $\text{\citep{sato2013planetary}}$& $1.797 \pm 0.052$ &  $2.661 \pm 0.335$ &  $ 11.03\pm 0.654$  & $ 2.779 \pm 0.075$ & Red Clump & 73.498$\pm 0.734$ & 6.036$\pm 0.131$ & 303.96 & $ 0.30$ & 0.38 &  & HD 120084 b&    $ 1373.9 \pm 128.6$ & $ 4.323 \pm 0.405$  & $ 4.513 \pm 0.263$\\ 
	{HD 29399}& 4845$\pm  52$ $\text{\citep{pezzotti2022coralie}}$& $ 1.067 \pm 0.024$ & $ 1.361 \pm 0.057$ &  $ 4.845 \pm 0.082$ & $ 3.202 \pm 0.023$ & RGB & 195.569 $\pm 0.707$ & 14.707 $\pm 0.058$ & 83.55 & $ 0.09$ & 0.15 & &  HD 29399 b &  $ 551.8 \pm 48.7$ &   $ 1.736 \pm 0.153$ &   $ 2.012 \pm 0.054$\\
	\enddata
	\tablecomments{Because the stellar mass calculated in this work through asteroseismology differs from the data provided in previous literature, it has altered the minimum mass and semi-major axis of the planet. The planetary data displayed in the table were derived solely from existing radial velocity data. The exact mass of HD 120084 b, however, was confirmed through additional astrometric measurements \citep{teng2023revisiting}, which are unrelated to the content of this work, and therefore is not presented. }
	\end{splitdeluxetable*}

	\section{HD 120084 - a Red Clump Host Star} \label{sec:120084}

	HD 120084 (TIC 284181945) is a red giant star hosting a confirmed exoplanet, HD 120084 b, which was discovered in 2013 through the radial velocity method \citep{sato2013planetary}. This planet has a long orbital period of 2142 days and follows an elliptical orbit with a high eccentricity, by combining radial velocity data with astrometric measurements, the orbital inclination of the planet has been determined, allowing for the calculation of its mass as 6.4 times that of Jupiter, indicating it is a giant planet \citep{teng2023revisiting}. From August 2019 to February 2024, TESS observed HD 120084 over 13 Sectors, accumulating a substantial amount of asteroseismic photometric data. After removing outliers and detrending the light curves, we computed its power spectrum, and then using \texttt{pySYD} to calculate the global asteroseismic parameters, we found the $\nu_{max}$ to be $73.498\pm 0.734\mu$Hz and the $\Delta\nu$ to be $6.036\pm 0.131\mu$Hz. Next, we used the effective temperature given in the literature, which is $4892 \pm 22 K$ \citep{sato2013planetary}, combined with asteroseismic parameters, to calculate the mass, radius, luminosity, and surface gravity of the star. The specific data are displayed in Table \ref{tab:para}. 

	To identify the $l=1$ mixed modes, we first converted the power spectrum into a SNR (Signal-to-Noise Ratio) spectrum, as detailed in Section \ref{subsec:preprocess}. Using observed value of asteroseismic parameters and $T_{eff}$ as the prior distribution, combined with the SNR spectrum, we inputted the data into the \texttt{PBjam} package. \texttt{PBjam} is able to automatically identify the components belonging to $\nu_{n,0}$ and $\nu_{n,2}$ in the spectrum. As shown in the bottom left corner of Figure A1 (see supplementary materials), the modes that are located in the middle part of the echelle diagram with a higher SNR are very likely to be the $l=1$ mixed modes. First, we filtered the power spectrum data between 40 and 110 $\mu Hz$, then removed the data between $\nu_{n,2} - 0.1\Delta\nu$ and $\nu_{n,0} + 0.1\Delta\nu$. Based on Section \ref{subsec:periodspacing}, we obtained $\Delta\Pi_{1}$ equal to $303.96 \pm 0.30$ seconds and q value equal to 0.38.

\begin{figure}[h!]
		
		\includegraphics[width=1.0\linewidth]{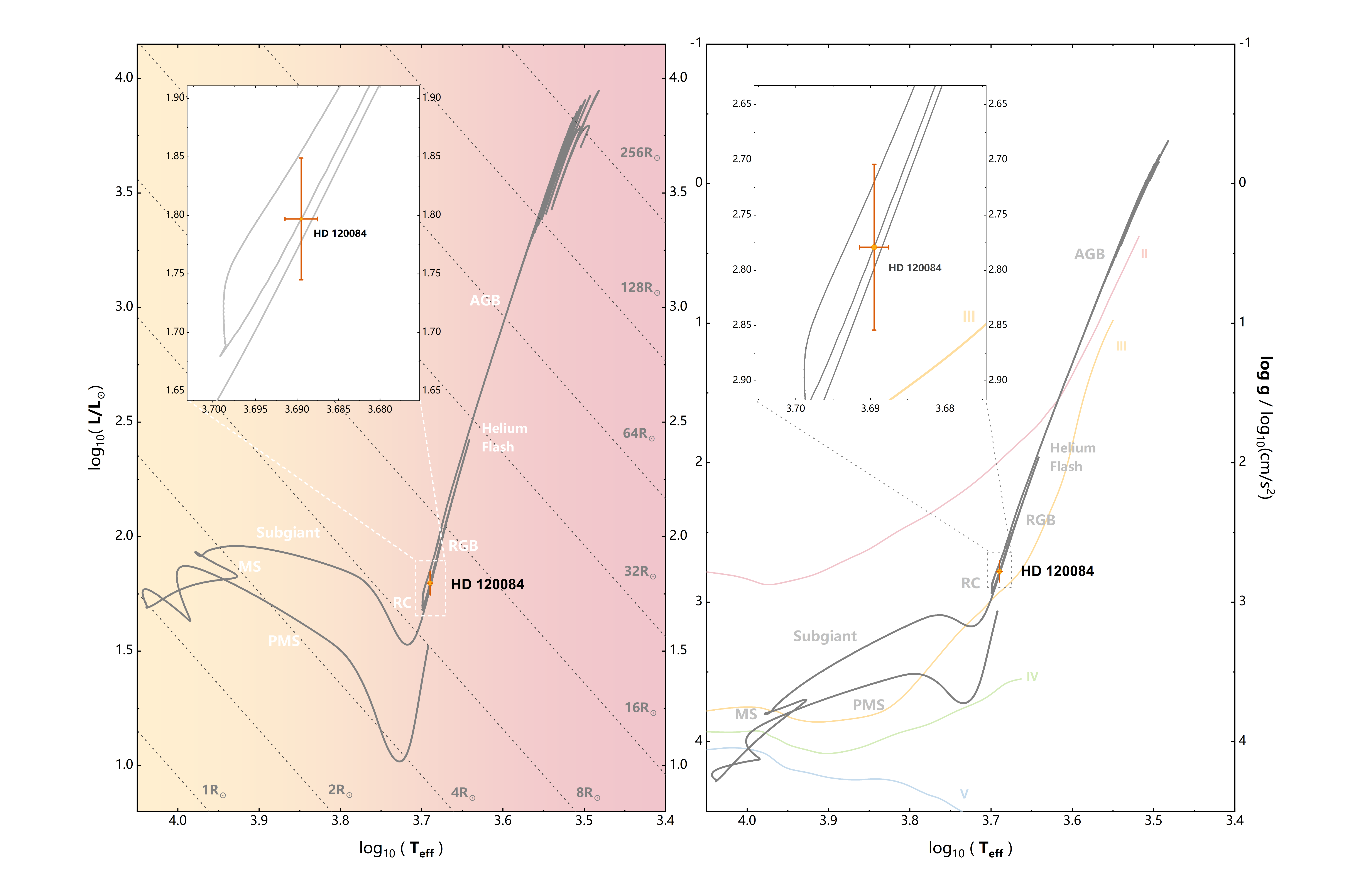}
		\centering
		\caption{\textbf{Left}: The effective temperature-luminosity Hertzsprung-Russell diagram of HD 120084, with the grey line representing the evolutionary track of HD 120084 simulated numerically by MESA, and the red data points and error bars derived from the data in this work. \textbf{Right}: The effective temperature-surface gravity Hertzsprung-Russell diagram of HD 120084. The orange line (III) and green line (IV) represents the giants and subgiant respectively, the red lines mark the positions of supergiants (II), the blue lines mark the positions of dwarfs (V) \citep{straivzys1981fundamental}.}
		\label{fig:MESA120084}

	\end{figure}

	The above analysis indicates that the $\Delta\Pi_{1}$ value of HD 120084 is consistent with the typical period spacing of dipole gravity modes in red clump giants undergoing helium core burning, which is between 280 and 320 seconds \citep[see][Fig. 9]{vrard2016period}. This suggests that HD 120084 is currently in the helium core burning phase.
	We used the \texttt{1M\_pre\_ms\_to\_wd} test suite case of MESA (r22.11.1) (Paxton et al. \citeyear{Paxton2011,2013ApJS..208....4P,2015ApJS..220...15P,2018ApJS..234...34P,2019ApJS..243...10P}) to simulate the evolutionary track of HD 120084, with an initial mass of 2.6719 $M_{\odot}$ (Within 0.1 $\sigma$ of the asteroseismic mass) and a metallicity of 0.0254 (Within 0.3 $\sigma$ of the metallicity in \citet{sato2013planetary}). As shown in Figure \ref{fig:MESA120084}, during the helium-core burning phase of this model, the model parameters are all within 0.1 sigma of the observed values of HD 120084, suggesting that HD 120084 can be considered a helium-core burning star. However, the hydrogen shell burning phase is also within 1 sigma of the observed values. This precisely illustrates that it is unlikely to distinguish between the evolutionary states before and after the helium flash based solely on the evolutionary model and stellar atmospheric parameters.
	Without an analysis of the mixed modes, it would not be possible to assertively determine the stage of the star through theoretical models. What's fascinating is the confirmed presence of a giant planet within this star system with a long orbital period. Whether this star engulfed any close-in planets during its past evolutionary stages will become the focus of the next phase of research. Future observations and studies will reveal more clues about the evolution of the system. Undoubtedly, HD 120084 is one of the vital and rare samples for studying the fate of planets.

	\section{HD 29399 - a Red Giant Branch Host Star} \label{sec:29399}
	HD 29399 (TIC 38828538) is a star in the red giant branch phase and has one confirmed exoplanet, HD 29399 b, which has an orbital period of 892.7 days. It was discovered using the radial velocity method \cite[]{pezzotti2022coralie}. In the literature where this exoplanet was first discovered \cite[]{pezzotti2022coralie}, data from TESS observations were also included for asteroseismic analysis. However, there was a lack of further analysis of $\nu_{n,1}$, and it did not provide direct evidence that HD 29399 is an RGB star. In this work, we used the same analysis method in Section \ref{subsec:periodspacing} to demonstrate that HD 29399 is an RGB star in the hydrogen shell burning phase. Additionally, since the work of \cite{pezzotti2022coralie}, TESS has accumulated more observational data, improving the signal-to-noise ratio.

	From August 2018 to September 2023, TESS observed HD 29399 across 32 sectors. Using the same method described in Section \ref{sec:120084}, we obtained the star's numax as $195.569\pm0.707$ $\mu Hz$ and deltanu as $14.707\pm0.058$ $\mu Hz$. Based on these findings, along with the effective temperature quoted from \cite{pezzotti2022coralie}, $4845\pm52$ K, the asteroseismic parameters of the star are listed in Table \ref{tab:para}. Similarly, after using the \texttt{PBjam} package to identify $\nu_{n,0}$ and $\nu_{n,2}$, we first filtered the spectrum data between 110 and 250 µHz, then removed the data between $\nu_{n,2} - 0.1\Delta\nu$ and $\nu_{n,0} + 0.3\Delta\nu$. Note that here, we removed more data beyond $\nu_{n,0}$ (the removal length was adjusted to 0.3 $\Delta\nu$) because visually identifiable $\nu_{n,3}$ modes appeared, and their mixed gravity modes differ from $\nu_{n,1}$ modes, thus requiring removal. We used the same method as described in Section \ref{sec:120084} to determine the $\Delta\Pi_{1}$ and $q$, which are $83.55\pm0.09$ seconds and 0.15, respectively. These values indicate that HD 29399 is a star at the hydrogen shell burning stage, during which the typical $\Delta\Pi_{1}$ value is between 65 and 90 seconds \citep[see][Fig. 9]{vrard2016period}.

	We used the \texttt{1M\_pre\_ms\_to\_wd} test suite case of MESA to simulate the evolutionary track of HD 29399, with an initial mass of 1.3668 $M_{\odot}$ (About 0.1 $\sigma$ of the asteroseismic mass) and a metallicity of 0.0249 (Within 1.5 $\sigma$ of the metallicity in \citet{pezzotti2022coralie}).
	The simulation results from MESA also show that HD 29399 is a Red Giant Branch (RGB) star within 0.1 $\sigma$ of the observed values for luminosity, effective temperature, and surface gravity, which is quite far from the red clump stage (see supplementary materials). The conclusions from various analyses are consistent with those of \cite{pezzotti2022coralie}, thus this example indeed substantiates the accuracy of asteroseismology in determining the evolutionary stage of stars.

	\section{Conlusion} \label{sec:conlusion}

	This study has conducted detailed asteroseismic analyses on two post-main-sequence stars, HD 120084 and HD 29399, known to host exoplanets, utilizing data from TESS. Through meticulous processing of the light curves and estimation of asteroseismic parameters, we have successfully identified their evolutionary states and accurately calculated the fundamental physical parameters of the stars. 
	Compared to the results given in the published papers, 
	$M_{HD 120084} = 2.39_{-0.30}^{+0.06}M_{\odot}$, 
	$R_{HD 120084} = 9.12_{-0.61}^{+0.65}R_{\odot}$ \citep{sato2013planetary}, 
	$M_{HD 29399} = 1.17\pm0.10 M_{\odot}$, 
	$R_{HD 29399} = 4.50\pm0.11 R_{\odot}$\citep{pezzotti2022coralie}, 
	the stellar parameters obtained through asteroseismology in this study (see Table \ref{tab:para}) can be considered consistent with the parameters in the papers within 2 $\sigma$.
	Consequently, the planetary parameters derived from these are also consistent. Therefore, it is feasible to determine the parameters of the planet using the asteroseismic data of the host star. The analysis of HD 120084 indicates that it is a star in the Red Clump phase, while HD 29399 is identified as being on the Red Giant Branch. Both examples demonstrate the power of asteroseismology as an effective tool for determining the evolutionary stage of stars, providing key insights into the fate of exoplanetary systems.

	Our findings highlight the importance of asteroseismology in contemporary astronomy, especially in exploring the interplay between stellar and planetary evolution. The planetary companion of HD 120084, with its long orbital period and high eccentricity, presents an important case for studying the potential engulfment of close-in planets during stellar evolution. On the other hand, the analysis of HD 29399 offers a detailed physical portrait of a star in the RGB phase, enhancing our understanding of this critical evolutionary stage.

	Overall, this work not only showcases the technical prowess of asteroseismology in analyzing stellar data but also opens new perspectives on understanding the interactions between stars and planets and their impact on the fate of planetary systems. Future observations and research will further unveil the dynamic evolutionary processes of these complex systems, providing us with valuable knowledge on the evolution of planetary systems in the universe.

	\section*{Acknowledgments}

	This work is supported by National Key R\&D Progrom of China (grant No:2022YFE0116800),the National Natural Science Foundation of China (No. 11933008 and No. 12103084), the basic research project of Yunnan Province (Grant No. 202301AT070352). This paper includes data collected by the TESS mission, which are publicly available from the Mikulski Archive for Space Telescopes (MAST)\footnote{https://mast.stsci.edu/}. Funding for the TESS mission is provided by the NASA Science Mission Directorate.

	Additionally, we are very grateful for the code and methodological support provided by Dr. Mathieu Vrard \footnote{vrard.1@osu.edu}.

	The data of MESA inlist is available on Zenodo under an open-source Creative Commons Attribution license: \dataset[doi:10.5281/zenodo.12702683]{https://doi.org/10.5281/zenodo.12702683}.

	\bibliography{sample631}{}

\begin{thebibliography}{}
\expandafter\ifx\csname natexlab\endcsname\relax\def\natexlab#1{#1}\fi
\providecommand{\url}[1]{\href{#1}{#1}}
\providecommand{\dodoi}[1]{doi:~\href{http://doi.org/#1}{\nolinkurl{#1}}}
\providecommand{\doeprint}[1]{\href{http://ascl.net/#1}{\nolinkurl{http://ascl.net/#1}}}
\providecommand{\doarXiv}[1]{\href{https://arxiv.org/abs/#1}{\nolinkurl{https://arxiv.org/abs/#1}}}

\bibitem[{Andersen {et~al.}(2019)Andersen, Pall{\'e}, Jessen-Hansen, Wang, Grundahl, Bedding, Cortes, Yu, Mathur, Gacia, {et~al.}}]{andersen2019oscillations}
Andersen, M.~F., Pall{\'e}, P., Jessen-Hansen, J., {et~al.} 2019, Astronomy \& Astrophysics, 623, L9

\bibitem[{Beck {et~al.}(2011)Beck, Bedding, Mosser, Stello, Garcia, Kallinger, Hekker, Elsworth, Frandsen, Carrier, {et~al.}}]{beck2011kepler}
Beck, P., Bedding, T.~R., Mosser, B., {et~al.} 2011, Science, 332, 205

\bibitem[{Bedding {et~al.}(2010)Bedding, Huber, Stello, Elsworth, Hekker, Kallinger, Mathur, Mosser, Preston, Ballot, {et~al.}}]{bedding2010solar}
Bedding, T.~R., Huber, D., Stello, D., {et~al.} 2010, The Astrophysical Journal Letters, 713, L176

\bibitem[{Bedding {et~al.}(2011)Bedding, Mosser, Huber, Montalb{\'a}n, Beck, Christensen-Dalsgaard, Elsworth, Garc{\'\i}a, Miglio, Stello, {et~al.}}]{bedding2011gravity}
Bedding, T.~R., Mosser, B., Huber, D., {et~al.} 2011, Nature, 471, 608

\bibitem[{Bowman(2020)}]{bowman2020asteroseismology}
Bowman, D.~M. 2020, Frontiers in Astronomy and Space Sciences, 7, 578584

\bibitem[{Campante {et~al.}(2019)Campante, Corsaro, Lund, Mosser, Serenelli, Veras, Adibekyan, Antia, Ball, Basu, {et~al.}}]{campante2019tess}
Campante, T.~L., Corsaro, E., Lund, M.~N., {et~al.} 2019, The Astrophysical Journal, 885, 31

\bibitem[{{Campante} {et~al.}(2019){Campante}, {Corsaro}, {Lund}, {Mosser}, {Serenelli}, {Veras}, {Adibekyan}, {Antia}, {Ball}, {Basu}, {Bedding}, {Bossini}, {Davies}, {Delgado Mena}, {Garc{\'\i}a}, {Handberg}, {Hon}, {Kane}, {Kawaler}, {Kuszlewicz}, {Lucas}, {Mathur}, {Nardetto}, {Nielsen}, {Pinsonneault}, {Reffert}, {Silva Aguirre}, {Stassun}, {Stello}, {Stock}, {Vrard}, {Y{\i}ld{\i}z}, {Chaplin}, {Huber}, {Bean}, {{\c{C}}elik Orhan}, {Cunha}, {Christensen-Dalsgaard}, {Kjeldsen}, {Metcalfe}, {Miglio}, {Monteiro}, {Nsamba}, {{\"O}rtel}, {Pereira}, {Sousa}, {Tsantaki}, \& {Turnbull}}]{2019ApJ...885...31C}
{Campante}, T.~L., {Corsaro}, E., {Lund}, M.~N., {et~al.} 2019, \apj, 885, 31, \dodoi{10.3847/1538-4357/ab44a8}

\bibitem[{Chaplin \& Miglio(2013)}]{chaplin2013asteroseismology}
Chaplin, W.~J., \& Miglio, A. 2013, Annual Review of Astronomy and Astrophysics, 51, 353

\bibitem[{{Chontos} {et~al.}(2022){Chontos}, {Huber}, {Sayeed}, \& {Yamsiri}}]{2022JOSS....7.3331C}
{Chontos}, A., {Huber}, D., {Sayeed}, M., \& {Yamsiri}, P. 2022, The Journal of Open Source Software, 7, 3331, \dodoi{10.21105/joss.03331}

\bibitem[{Cunha {et~al.}(2019)Cunha, Avelino, Christensen-Dalsgaard, Stello, Vrard, Jiang, \& Mosser}]{cunha2019analytical}
Cunha, M., Avelino, P., Christensen-Dalsgaard, J., {et~al.} 2019, Monthly Notices of the Royal Astronomical Society, 490, 909

\bibitem[{{Deheuvels} {et~al.}(2015){Deheuvels}, {Ballot}, {Beck}, {Mosser}, {{\O}stensen}, {Garc{\'\i}a}, \& {Goupil}}]{2015A&A...580A..96D}
{Deheuvels}, S., {Ballot}, J., {Beck}, P.~G., {et~al.} 2015, \aap, 580, A96, \dodoi{10.1051/0004-6361/201526449}

\bibitem[{{Feng} {et~al.}(2022){Feng}, {Butler}, {Vogt}, {Clement}, {Tinney}, {Cui}, {Aizawa}, {Jones}, {Bailey}, {Burt}, {Carter}, {Crane}, {Flammini Dotti}, {Holden}, {Ma}, {Ogihara}, {Oppenheimer}, {O'Toole}, {Shectman}, {Wittenmyer}, {Wang}, {Wright}, \& {Xuan}}]{2022ApJS..262...21F}
{Feng}, F., {Butler}, R.~P., {Vogt}, S.~S., {et~al.} 2022, \apjs, 262, 21, \dodoi{10.3847/1538-4365/ac7e57}

\bibitem[{{Garc{\'\i}a} \& {Ballot}(2019)}]{2019LRSP...16....4G}
{Garc{\'\i}a}, R.~A., \& {Ballot}, J. 2019, Living Reviews in Solar Physics, 16, 4, \dodoi{10.1007/s41116-019-0020-1}

\bibitem[{Giacobbe {et~al.}(2021)Giacobbe, Brogi, Gandhi, Cubillos, Bonomo, Sozzetti, Fossati, Guilluy, Carleo, Rainer, {et~al.}}]{giacobbe2021five}
Giacobbe, P., Brogi, M., Gandhi, S., {et~al.} 2021, Nature, 592, 205

\bibitem[{Guo(2024)}]{guo2024characterization}
Guo, J. 2024, Nature Astronomy, 1

\bibitem[{Hon {et~al.}(2023)Hon, Huber, Rui, Fuller, Veras, Kuszlewicz, Kochukhov, Stokholm, R{\o}rsted, Y{\i}ld{\i}z, {et~al.}}]{hon2023close}
Hon, M., Huber, D., Rui, N.~Z., {et~al.} 2023, Nature, 618, 917

\bibitem[{{Jackiewicz}(2021)}]{2021FrASS...7..102J}
{Jackiewicz}, J. 2021, Frontiers in Astronomy and Space Sciences, 7, 102, \dodoi{10.3389/fspas.2020.595017}

\bibitem[{Jenkins {et~al.}(2016)Jenkins, Twicken, McCauliff, Campbell, Sanderfer, Lung, Mansouri-Samani, Girouard, Tenenbaum, Klaus, {et~al.}}]{jenkins2016tess}
Jenkins, J.~M., Twicken, J.~D., McCauliff, S., {et~al.} 2016, in Software and Cyberinfrastructure for Astronomy IV, Vol. 9913, SPIE, 1232--1251

\bibitem[{Jontof-Hutter(2019)}]{jontof2019compositional}
Jontof-Hutter, D. 2019, Annual Review of Earth and Planetary Sciences, 47, 141

\bibitem[{{Kjeldsen} \& {Bedding}(1995)}]{1995A&A...293...87K}
{Kjeldsen}, H., \& {Bedding}, T.~R. 1995, \aap, 293, 87, \dodoi{10.48550/arXiv.astro-ph/9403015}

\bibitem[{Konacki {et~al.}(2003)Konacki, Torres, Jha, \& Sasselov}]{konacki2003extrasolar}
Konacki, M., Torres, G., Jha, S., \& Sasselov, D.~D. 2003, Nature, 421, 507

\bibitem[{{Lightkurve Collaboration} {et~al.}(2018){Lightkurve Collaboration}, {Cardoso}, {Hedges}, {Gully-Santiago}, {Saunders}, {Cody}, {Barclay}, {Hall}, {Sagear}, {Turtelboom}, {Zhang}, {Tzanidakis}, {Mighell}, {Coughlin}, {Bell}, {Berta-Thompson}, {Williams}, {Dotson}, \& {Barentsen}}]{2018ascl.soft12013L}
{Lightkurve Collaboration}, {Cardoso}, J.~V.~d.~M., {Hedges}, C., {et~al.} 2018, {Lightkurve: Kepler and TESS time series analysis in Python}, Astrophysics Source Code Library.
\newblock \doeprint{1812.013}

\bibitem[{{Lin} {et~al.}(2024){Lin}, {Qian}, {Zhu}, {Liao}, \& {Li}}]{2024AJ....168...27L}
{Lin}, W.-X., {Qian}, S.-B., {Zhu}, L.-Y., {Liao}, W.-P., \& {Li}, F.-X. 2024, \aj, 168, 27, \dodoi{10.3847/1538-3881/ad4ffc}

\bibitem[{Lomb(1976)}]{lomb1976least}
Lomb, N.~R. 1976, Astrophysics and space science, 39, 447

\bibitem[{{Lundkvist} {et~al.}(2018){Lundkvist}, {Huber}, {Silva Aguirre}, \& {Chaplin}}]{2018arXiv180402214L}
{Lundkvist}, M.~S., {Huber}, D., {Silva Aguirre}, V., \& {Chaplin}, W.~J. 2018, arXiv e-prints, arXiv:1804.02214, \dodoi{10.48550/arXiv.1804.02214}

\bibitem[{Mayor \& Queloz(1995)}]{mayor1995jupiter}
Mayor, M., \& Queloz, D. 1995, nature, 378, 355

\bibitem[{{Mosser} {et~al.}(2015){Mosser}, {Vrard}, {Belkacem}, {Deheuvels}, \& {Goupil}}]{2015A&A...584A..50M}
{Mosser}, B., {Vrard}, M., {Belkacem}, K., {Deheuvels}, S., \& {Goupil}, M.~J. 2015, \aap, 584, A50, \dodoi{10.1051/0004-6361/201527075}

\bibitem[{Mosser {et~al.}(2012)Mosser, Goupil, Belkacem, Michel, Stello, Marques, Elsworth, Barban, Beck, Bedding, {et~al.}}]{mosser2012probing}
Mosser, B., Goupil, M., Belkacem, K., {et~al.} 2012, Astronomy \& Astrophysics, 540, A143

\bibitem[{Mulders(2018)}]{mulders2018planet}
Mulders, G.~D. 2018, arXiv preprint arXiv:1805.00023

\bibitem[{{NASA Exoplanet Archive}(2024{\natexlab{a}})}]{STELLARHOSTS}
{NASA Exoplanet Archive}. 2024{\natexlab{a}}, Stellar Parameters Table,  NExScI-Caltech/IPAC, \dodoi{10.26133/NEA40}

\bibitem[{{NASA Exoplanet Archive}(2024{\natexlab{b}})}]{ps}
---. 2024{\natexlab{b}}, Planetary Systems,  NExScI-Caltech/IPAC, \dodoi{10.26133/NEA12}

\bibitem[{{Nielsen} {et~al.}(2021){Nielsen}, {Davies}, {Ball}, {Lyttle}, {Li}, {Hall}, {Chaplin}, {Gaulme}, {Carboneau}, {Ong}, {Garc{\'\i}a}, {Mosser}, {Roxburgh}, {Corsaro}, {Benomar}, {Moya}, \& {Lund}}]{2021AJ....161...62N}
{Nielsen}, M.~B., {Davies}, G.~R., {Ball}, W.~H., {et~al.} 2021, \aj, 161, 62, \dodoi{10.3847/1538-3881/abcd39}

\bibitem[{{Paxton} {et~al.}(2011){Paxton}, {Bildsten}, {Dotter}, {Herwig}, {Lesaffre}, \& {Timmes}}]{Paxton2011}
{Paxton}, B., {Bildsten}, L., {Dotter}, A., {et~al.} 2011, \apjs, 192, 3, \dodoi{10.1088/0067-0049/192/1/3}

\bibitem[{{Paxton} {et~al.}(2013){Paxton}, {Cantiello}, {Arras}, {Bildsten}, {Brown}, {Dotter}, {Mankovich}, {Montgomery}, {Stello}, {Timmes}, \& {Townsend}}]{2013ApJS..208....4P}
{Paxton}, B., {Cantiello}, M., {Arras}, P., {et~al.} 2013, \apjs, 208, 4, \dodoi{10.1088/0067-0049/208/1/4}

\bibitem[{{Paxton} {et~al.}(2015){Paxton}, {Marchant}, {Schwab}, {Bauer}, {Bildsten}, {Cantiello}, {Dessart}, {Farmer}, {Hu}, {Langer}, {Townsend}, {Townsley}, \& {Timmes}}]{2015ApJS..220...15P}
{Paxton}, B., {Marchant}, P., {Schwab}, J., {et~al.} 2015, \apjs, 220, 15, \dodoi{10.1088/0067-0049/220/1/15}

\bibitem[{{Paxton} {et~al.}(2018){Paxton}, {Schwab}, {Bauer}, {Bildsten}, {Blinnikov}, {Duffell}, {Farmer}, {Goldberg}, {Marchant}, {Sorokina}, {Thoul}, {Townsend}, \& {Timmes}}]{2018ApJS..234...34P}
{Paxton}, B., {Schwab}, J., {Bauer}, E.~B., {et~al.} 2018, \apjs, 234, 34, \dodoi{10.3847/1538-4365/aaa5a8}

\bibitem[{{Paxton} {et~al.}(2019){Paxton}, {Smolec}, {Schwab}, {Gautschy}, {Bildsten}, {Cantiello}, {Dotter}, {Farmer}, {Goldberg}, {Jermyn}, {Kanbur}, {Marchant}, {Thoul}, {Townsend}, {Wolf}, {Zhang}, \& {Timmes}}]{2019ApJS..243...10P}
{Paxton}, B., {Smolec}, R., {Schwab}, J., {et~al.} 2019, \apjs, 243, 10, \dodoi{10.3847/1538-4365/ab2241}

\bibitem[{Pezzotti {et~al.}(2022)Pezzotti, Ottoni, Buldgen, Lyttle, Eggenberger, Udry, S{\'e}gransan, Mayor, Lovis, Marmier, {et~al.}}]{pezzotti2022coralie}
Pezzotti, C., Ottoni, G., Buldgen, G., {et~al.} 2022, Astronomy \& Astrophysics, 657, A89

\bibitem[{Pr{\v{s}}a {et~al.}(2016)Pr{\v{s}}a, Harmanec, Torres, Mamajek, Asplund, Capitaine, Christensen-Dalsgaard, Depagne, Haberreiter, Hekker, {et~al.}}]{prvsa2016nominal}
Pr{\v{s}}a, A., Harmanec, P., Torres, G., {et~al.} 2016, The Astronomical Journal, 152, 41

\bibitem[{Sato {et~al.}(2013)Sato, Omiya, Harakawa, Liu, Izumiura, Kambe, Takeda, Yoshida, Itoh, Ando, {et~al.}}]{sato2013planetary}
Sato, B., Omiya, M., Harakawa, H., {et~al.} 2013, Publications of the Astronomical Society of Japan, 65, 85

\bibitem[{Scargle(1982)}]{scargle1982studies}
Scargle, J.~D. 1982, Astrophysical Journal, Part 1, vol. 263, Dec. 15, 1982, p. 835-853., 263, 835

\bibitem[{Sharma {et~al.}(2016)Sharma, Stello, Bland-Hawthorn, Huber, \& Bedding}]{sharma2016stellar}
Sharma, S., Stello, D., Bland-Hawthorn, J., Huber, D., \& Bedding, T.~R. 2016, The Astrophysical Journal, 822, 15

\bibitem[{Southworth(2021)}]{southworth2021space}
Southworth, J. 2021, Universe, 7, 369

\bibitem[{Stello \& Sharma(2023)}]{stello2023extension}
Stello, D., \& Sharma, S. 2023, arXiv preprint arXiv:2305.03221

\bibitem[{Stello {et~al.}(2013)Stello, Huber, Bedding, Benomar, Bildsten, Elsworth, Gilliland, Mosser, Paxton, \& White}]{stello2013asteroseismic}
Stello, D., Huber, D., Bedding, T.~R., {et~al.} 2013, The Astrophysical Journal Letters, 765, L41

\bibitem[{Strai{\v{z}}ys \& Kuriliene(1981)}]{straivzys1981fundamental}
Strai{\v{z}}ys, V., \& Kuriliene, G. 1981, Astrophysics and Space Science, 80, 353

\bibitem[{Teng {et~al.}(2023)Teng, Sato, Kuzuhara, Takarada, Omiya, Harakawa, Izumiura, Kambe, Yilmaz, Bikmaev, {et~al.}}]{teng2023revisiting}
Teng, H.-Y., Sato, B., Kuzuhara, M., {et~al.} 2023, Publications of the Astronomical Society of Japan, 75, 1030

\bibitem[{{Teng} {et~al.}(2023){Teng}, {Sato}, {Kuzuhara}, {Takarada}, {Omiya}, {Harakawa}, {Izumiura}, {Kambe}, {Yilmaz}, {Bikmaev}, {Selam}, {Brandt}, {Xiao}, {Yoshida}, {Itoh}, {Ando}, {Kokubo}, \& {Ida}}]{2023PASJ...75.1030T}
{Teng}, H.-Y., {Sato}, B., {Kuzuhara}, M., {et~al.} 2023, \pasj, 75, 1030, \dodoi{10.1093/pasj/psad056}

\bibitem[{{Unno} {et~al.}(1989){Unno}, {Osaki}, {Ando}, {Saio}, \& {Shibahashi}}]{1989nos..book.....U}
{Unno}, W., {Osaki}, Y., {Ando}, H., {Saio}, H., \& {Shibahashi}, H. 1989, {Nonradial oscillations of stars}

\bibitem[{{Unno} {et~al.}(1979){Unno}, {Osaki}, {Ando}, \& {Shibahashi}}]{1979nos..book.....U}
{Unno}, W., {Osaki}, Y., {Ando}, H., \& {Shibahashi}, H. 1979, {Nonradial oscillations of stars}

\bibitem[{Vrard {et~al.}(2016)Vrard, Mosser, \& Samadi}]{vrard2016period}
Vrard, M., Mosser, B., \& Samadi, R. 2016, Astronomy \& Astrophysics, 588, A87

\bibitem[{Winn \& Fabrycky(2015)}]{winn2015occurrence}
Winn, J.~N., \& Fabrycky, D.~C. 2015, Annual Review of Astronomy and Astrophysics, 53, 409

\bibitem[{{Xiao} {et~al.}(2023){Xiao}, {Liu}, {Teng}, {Wang}, {Brandt}, {Zhao}, {Zhao}, {Zhai}, \& {Gao}}]{2023RAA....23e5022X}
{Xiao}, G.-Y., {Liu}, Y.-J., {Teng}, H.-Y., {et~al.} 2023, Research in Astronomy and Astrophysics, 23, 055022, \dodoi{10.1088/1674-4527/accb7e}

\end{thebibliography}
	\bibliographystyle{aasjournal}

	\section{Appendix}\label{sec:Appendix}
	This is a supplementary materials to the paper, and it contains one table and four figures. Table \ref{tab:freqs} lists the $l=0$ and $l=2$ modes identified for the two host stars using the $PBjam$ package, with eight modes each. Figure \ref{fig:HD 120084 analysis} is a composite of a portion of HD 120084's light curve, the $pySYD$ output, and the echelle diagram produced after $PBjam$ mode identification. Figure \ref{fig:HD 29399 analysis} is a composite of a portion of HD 29399's light curve, the $pySYD$ output, and the echelle diagram produced after $PBjam$ mode identification. Figure \ref{fig:MESA29399} is the MESA theoretical model diagram for HD 29399. Figures \ref{fig:HD 120084 corner} and \ref{fig:HD 29399 corner} are the corner diagram of asymptotic fit parameters plots output by $PBjam$.

	\renewcommand\thetable{A1}
	\begin{table}[h!]
		\centering
		\caption{\textbf{$\nu_{n,0}$ and $\nu_{n,2}$ of 2 host stars}}
		\label{tab:freqs}
		\tabletypesize{\scriptsize}
		\tablewidth{0pt}
		\renewcommand{\arraystretch}{1.5}
		\begin{tabular}{cccc}
		  \hline
		  \hline
			\multicolumn{2}{c}{HD 120084} & \multicolumn{2}{c}{HD 29399} \\
			$\nu_{n,0}$ [$\mu Hz$] & $\nu_{n,2}$ [$\mu Hz$] & $\nu_{n,0}$ [$\mu Hz$] & $\nu_{n,2}$ [$\mu Hz$] \\
		  \hline
		  51.949$\pm0.092$ & 51.369$\pm0.165$ & 139.655$\pm0.045$ & 137.643$\pm0.453$ \\
		  58.287$\pm0.033$ & 57.315$\pm0.056$ & 154.053$\pm0.023$ & 152.209$\pm0.045$ \\
		  64.418$\pm0.040$ & 63.574$\pm0.091$ & 168.631$\pm0.011$ & 166.676$\pm0.031$ \\
		  70.930$\pm0.019$ & 70.083$\pm0.034$ & 183.545$\pm0.009$ & 181.728$\pm0.010$ \\
		  77.347$\pm0.059$ & 76.436$\pm0.024$ & 198.356$\pm0.009$ & 196.491$\pm0.015$ \\
		  83.759$\pm0.038$ & 83.082$\pm0.038$ & 213.259$\pm0.021$ & 211.423$\pm0.012$ \\
		  90.378$\pm0.043$ & 89.390$\pm0.046$ & 228.523$\pm0.018$ & 226.709$\pm0.021$ \\
		  96.838$\pm0.067$ & 96.128$\pm0.167$ & 243.598$\pm0.091$ & 241.460$\pm0.262$ \\
		  \hline
		\end{tabular}
	\end{table}
	
	\renewcommand\thetable{A2}
	\begin{table}[h!]
		\centering
		\caption{Host star parameters in the published paper}
		\label{tab:paras_from_paper}
		\tabletypesize{\scriptsize}
		\tablewidth{0pt}
		\renewcommand{\arraystretch}{1.5}
		\begin{tabular}{ccccccc}
		\hline
		\hline
			\colhead{Host Star} & \colhead{Source} & \colhead{$T_{eff}$} & \colhead{log $g$} & \colhead{$M_{\star}$} & \colhead{$R_{\star}$} & \colhead{[Fe/H]}\\
			&  & (K) & ($\log_{10}{\mathrm{(cm/s^2)}} $) & ($M_{\odot}$) & ($R_{\odot}$) & (dex)\\
		\hline
		HD 120084 & \cite[]{2023PASJ...75.1030T} & 4897 & $2.74_{-0.09}^{+0.07}$ & $2.16_{-0.27}^{+0.28}$& $10.37_{-0.80}^{+0.82}$ & $0.10_{-0.10}^{+0.08}$ \\
			& \cite[]{2023RAA....23e5022X}& $4892\pm22$ & $2.71\pm0.08$ & $2.15\pm0.21$ & \nodata & $0.09\pm0.05$ \\
			& \cite[]{2022ApJS..262...21F} &\nodata &\nodata & $2.504\pm0.240$ &\nodata &\nodata\\
			&\cite[]{sato2013planetary} &$4892\pm22$ &$2.71\pm0.08$ &$2.39_{-0.30}^{+0.06}$ &$9.12_{-0.61}^{+0.65}$ &  $0.09\pm0.05$\\
		HD 29399 & \cite[]{pezzotti2022coralie}& $4845\pm52$&$3.25\pm0.13$ & $1.17\pm0.10$ & $4.50\pm0.11$& $0.14\pm0.03$ \\
		\hline
		\end{tabular}
	\end{table}

	\renewcommand\thefigure{A1}
	\begin{figure}[h!]

		\includegraphics[width=1.0\linewidth]{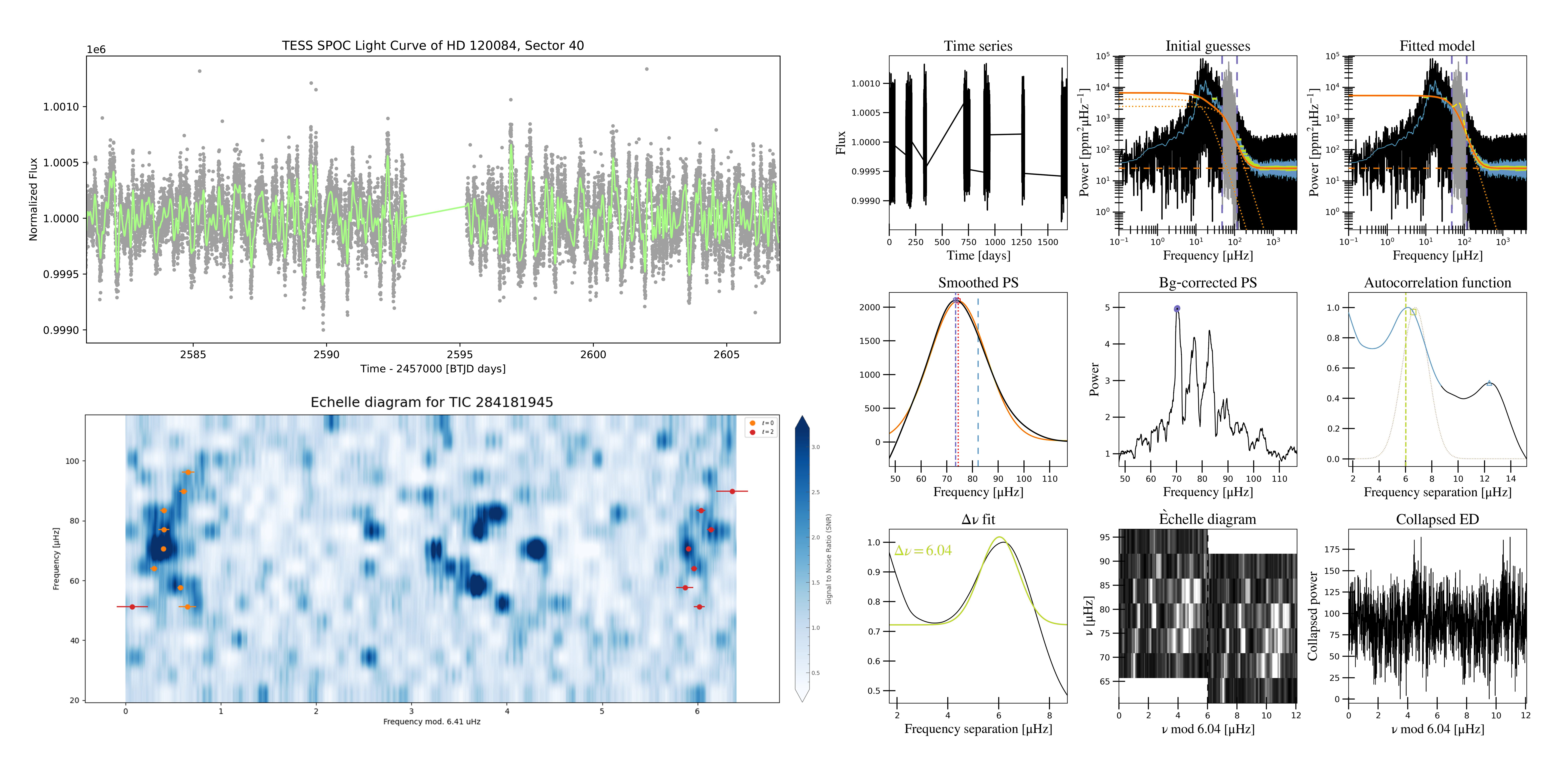}
		\centering
		\caption{\textbf{Top left}: The display of the light curve for Sector 40 of HD 120084 is described with gray dots representing the original data and a green line showing the smoothed trend of variation. \textbf{Right}: Display of the asteroseismic parameters obtained after analyzing the power spectrum with the $pySYD$ package, Detailed descriptions of each figure can be found at this \href{https://pysyd.readthedocs.io/en/latest/library/output.html}{website}. \textbf{Bottom left}: When the frequency separation is 6.41$\mu$Hz, the echelle diagram shows orange and red dots representing $\nu_{n,0}$ and $\nu_{n,2}$ as identified by PBjam. It is important to note that this frequency separation is not equivalent to the measured value of observed $\Delta\nu$.} 
		\label{fig:HD 120084 analysis}

	\end{figure}

	\renewcommand\thefigure{B1}
	\begin{figure}[h!]

		\includegraphics[width=1.0\linewidth]{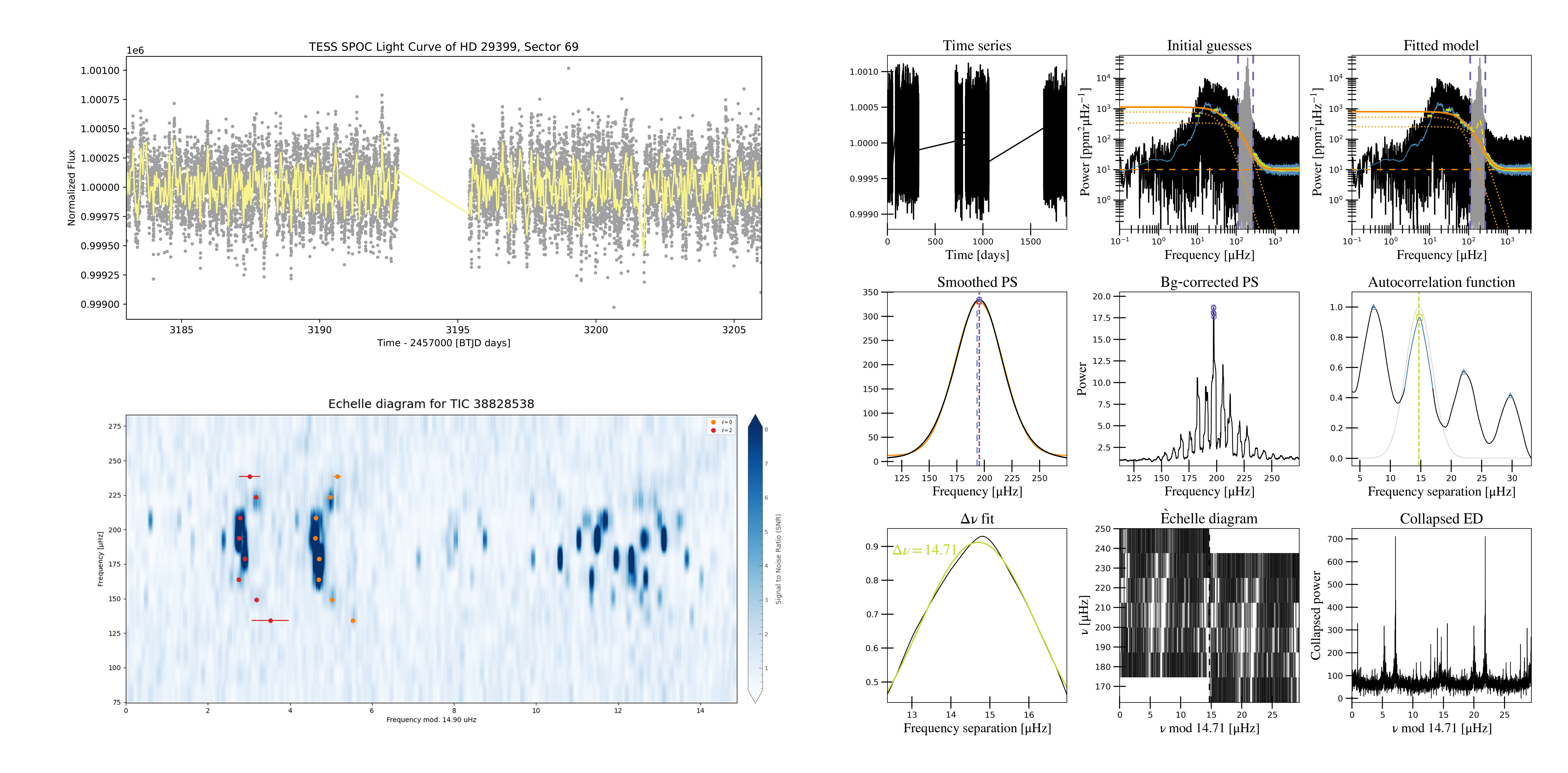}
		\centering
		\caption{\textbf{Top left}: The display of the light curve for Sector 40 of HD 29399 is described with gray dots representing the original data and a yellow line showing the smoothed trend of variation. \textbf{Right}: Display of the asteroseismic parameters obtained after analyzing the power spectrum with the $pySYD$ package, Detailed descriptions of each figure can be found at this \href{https://pysyd.readthedocs.io/en/latest/library/output.html}{website}. \textbf{Bottom left}: When the frequency separation is 6.41$\mu$Hz, the echelle diagram shows orange and red dots representing $\nu_{n,0}$ and $\nu_{n,2}$ as identified by PBjam. It is important to note that this frequency separation is not equivalent to the measured value of observed $\Delta\nu$.} 
		\label{fig:HD 29399 analysis}
	\end{figure}
	
	\renewcommand\thefigure{B2}
	\begin{figure}[h!]
	
		\includegraphics[width=1.0\linewidth]{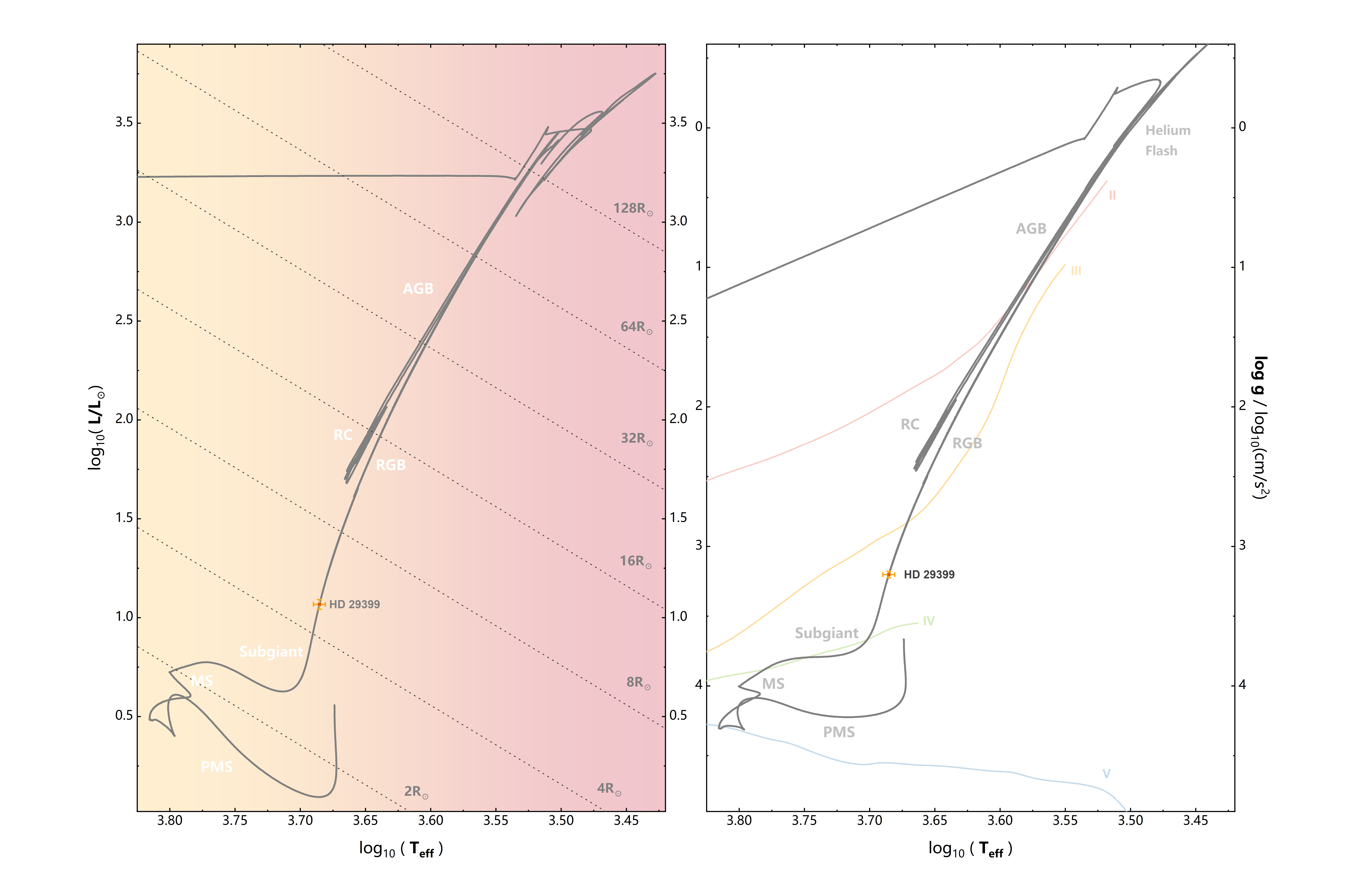}
		\centering
		\caption{\textbf{Left}: The effective temperature-luminosity Hertzsprung-Russell diagram of HD 29399, with the grey line representing the evolutionary track of HD 29399 simulated numerically by MESA, and the red data points and error bars derived from the data in this work. \textbf{Right}: The effective temperature-surface gravity Hertzsprung-Russell diagram of HD 29399} 
		\label{fig:MESA29399}
	
	\end{figure}

	\renewcommand\thefigure{C1}
	\begin{figure}[h!]

		\includegraphics[width=0.95\linewidth]{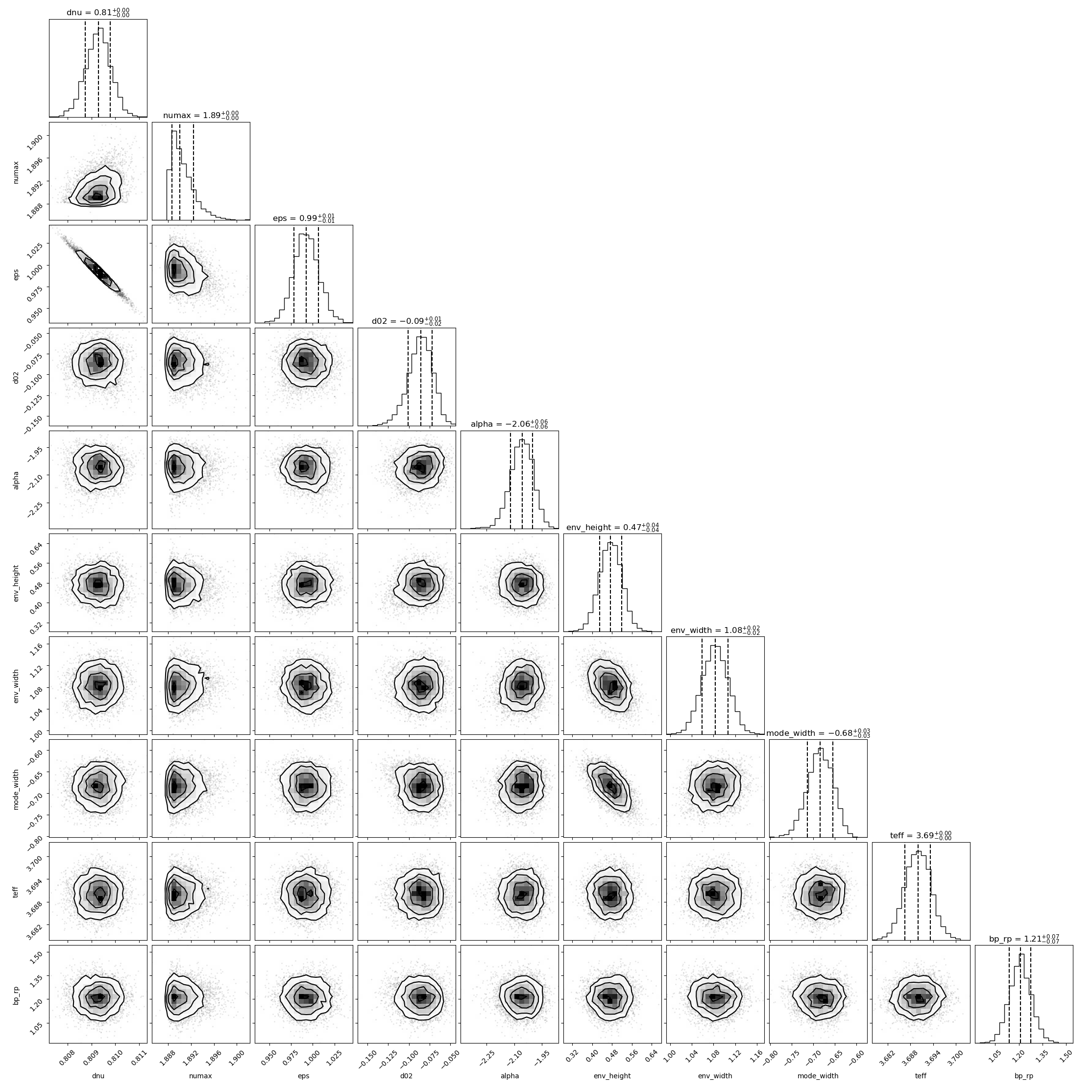}
		\centering
		\caption{HD 120084 corner diagram of asymptotic fit parameters.
		\label{fig:HD 120084 corner}} 
	\end{figure}

	\renewcommand\thefigure{C2}
	\begin{figure}[h!]

		\includegraphics[width=0.95\linewidth]{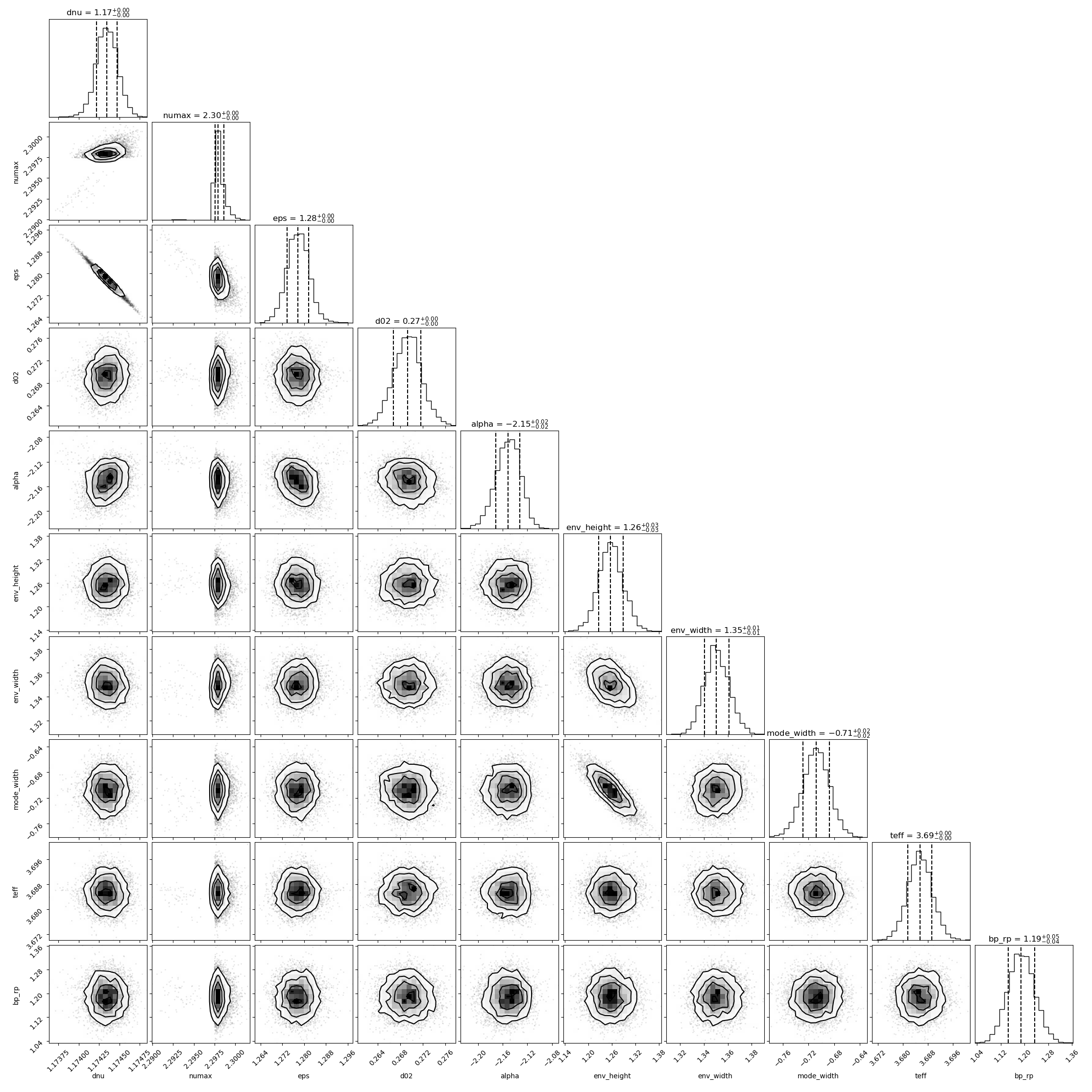}
		\centering
		\caption{HD 29399 corner diagram of asymptotic fit parameters.
		\label{fig:HD 29399 corner}} 
	\end{figure}

	\end{document}